\documentclass{article}
\usepackage{spconf,amsmath,graphicx}
\usepackage{epstopdf}
\usepackage{amsfonts,bm,booktabs,multirow,color}
\usepackage{subfigure}
\usepackage{url}

\title{Speech enhancement aided end-to-end multi-task learning for voice activity detection}
\name{Xu Tan, Xiao-Lei Zhang}
\address{CIAIC, School of Marine Science and Technology, Northwestern Polytechnical University, China}
\begin{document}
%
\maketitle
\begin{abstract}
	Robust voice activity detection (VAD) is a challenging task in low signal-to-noise (SNR) environments. Recent studies show that speech enhancement is helpful to VAD, but the performance improvement is limited. To address this issue, here we propose a speech enhancement aided end-to-end multi-task model for VAD. The model has two decoders, one for speech enhancement and the other for VAD. The two decoders share the same encoder and speech separation network. Unlike the direct thought that takes two separated objectives for VAD and speech enhancement respectively, here  we propose a new joint optimization objective---VAD-masked scale-invariant source-to-distortion ratio (mSI-SDR). mSI-SDR uses VAD information to mask the output of the speech enhancement decoder in the training process. It makes the VAD and speech enhancement tasks jointly optimized not only at the shared encoder and separation network, but also at the objective level. It also satisfies real-time working requirement theoretically.
Experimental results show that the multi-task method significantly outperforms its single-task VAD counterpart. Moreover, mSI-SDR outperforms SI-SDR in the same multi-task setting.
\end{abstract}
\section{Introduction}
\label{sec:intro}

Voice activity detection (VAD) aims to differentiate speech segments from noise segments in an audio recording. It is an important front-end for many speech-related applications, such as speech recognition and speaker recognition.
In recent years, deep learning based VAD have brought significant performance improvement \cite{zhang2012deep,hughes2013recurrent,thomas2014analyzing,zhang2015boosting,kim2018multi,KIM2018,chang2018,Fernando2020}. Particulary, the end-to-end VAD, which takes time-domain signals directly into deep networks, is a recent research trend\cite{Zazo2016,Ariav2019,Yu2020}.

\begin{figure}[t]
	
	\centering
	\centerline{\includegraphics[width=8.5cm]{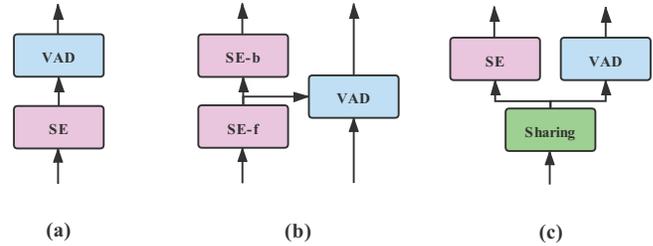}}
	\caption{Three typical architectures of speech enhancement aided VAD in literature.}
	\label{fig:jt}
\end{figure}

Although deep learning based VAD has shown its effectiveness, it is of long-time interests that how to further improve its performance in low signal-to-noise ratio (SNR) environments. A single VAD seems hard to meet the requirement. A natural thought is to bring speech enhancement (SE) into VAD. Several previous works have pursued this direction. The earliest method \cite{zhang2013denoising} uses a deep-learning-based speech enhancement network to initialize VAD. In \cite{Wang2015}, the authors uses a speech enhancement network to first denoise speech, and then uses the denoised speech as the input of VAD, where the enhancement network and VAD are jointly fine-tuned (Fig. \ref{fig:jt}a). Similar ideas can be found in \cite{Lin2019} too.

%

\begin{figure*}[t]
	
	\centering
	\centerline{\includegraphics[width=14cm]{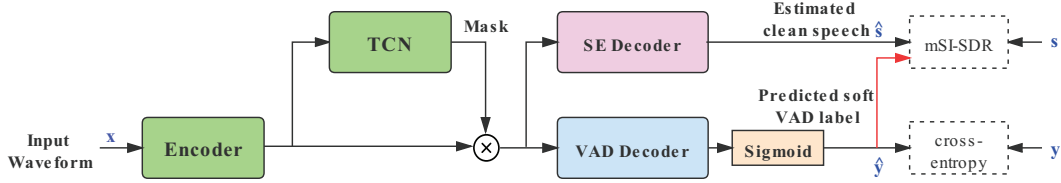}}
	\caption{Structure of the proposed end-to-end multi-task model. The red line denotes important information flow in the objective.}
	\label{fig:model}
\end{figure*}

Later on, it is observed that using the enhancement result as the input of VAD may do harm to VAD when the performance of the SE module is poor \cite{Xu2019}. Based on the observations, several work uses advanced speech enhancement methods to extract denoised features for VAD (Fig. \ref{fig:jt}b).
Lee \textit{et al.} \cite{Lee2020} used U-Net to estimate clean speech spectra and noise spectra simultaneously, and then used the enhanced speech spectrogram to conduct VAD directly by thresholding. Jung \textit{et al.} \cite{Jung2018} used the output and latent variable of a denoising variational autoencoder-based SE module as the input of VAD. Xu \textit{et al.} \cite{Xu2019} concatenated the noisy acoustic feature and an enhanced acoustic feature extracted from a convolutional-recurrent-network-based SE as the input of a residual-convolutional neural-network-based VAD.

Besides, Zhuang \textit{et al.} \cite{Zhuang2017} proposed multi-objective networks to jointly train SE and VAD for boosting both of their performance (Fig. \ref{fig:jt}c), where VAD and SE share the same network and have different loss functions. However, the performance improvement of VAD is limited. Here, we believe that the joint training strategy is promising, it is just unexplored deeply yet.

In this paper, we propose an end-to-end multi-task joint training model to improve the performance of VAD in adverse acoustic environments. Specifically, we employ Conv-TasNet \cite{Luo2019} as the backbone network. Then, we make SE and VAD share the same encoder and temporal convolutional network (TCN). Finally, we use two decoders for generating enhanced speech and speech likelihood ratios respectively. The novelties of the method are as follows

 \begin{itemize}
   \item To our knowledge, we propose the first end-to-end multitask model for VAD, where SE is used as an auxiliary task.
   \item We propose a novel loss function, named VAD-masked scale-invariant source-to-distortion ratio (mSI-SDR), at the SE decoder. It uses the the ground-truth and predicted VAD labels to mask the speech enhancement output. It makes the network structure different from the three classes of models in Fig. \ref{fig:jt}.
 \end{itemize}
 Besides, the proposed method also inherits the merit of low latency from Conv-TasNet. Experimental results demonstrate the effectiveness of the proposed end-to-end multi-task model as well as the advantage of the proposed mSI-SDR objective.

\section{End-to-end multi-task model with mSI-SDR}
\label{sec:model}

\subsection{Notation}
\label{ssec:review}

Given an audio signal of $T$ samples, denoted as $\mathbf{x} \in \mathbb{R}^{1 \times T}$, which is a mixture of clean speech $\mathbf{s}$ and noise $\mathbf{n}$, i.e. $\mathbf{x}=\mathbf{s}+\mathbf{n}$. Suppose $\mathbf{x}$ can be partitioned into $N$ frames. Usually, we transform the time-domain representation into a time-frequency representation $\{\mathbf{w}_i\}_{i=1}^N$. VAD first generates a soft prediction of $\mathbf{w}_{i}$, denoted as $\hat{y}_i$, and then compares $\hat{y}_i$ with a decision threshold for generating a hard decision, where $i$ denotes the $i$-th frame and $\hat{y}_i\in [0,1]$ is a soft prediction of the ground-truth label $y_i\in\{0,1\}$.
Speech enhancement aims to generate an estimate of $\mathbf{s}$, denoted as $\hat{\mathbf{s}}$, from $\mathbf{x}$.

\subsection{Network architecture}
\label{ssec:archi}

As shown in Fig. \ref{fig:model}, the proposed end-to-end multi-task model conducts speech enhancement and VAD simultaneously. It follows the architecture of Conv-TasNet \cite{Luo2019}, which contains three parts---an encoder, a separation network, and two decoders. The two tasks share the same encoder and separation network. Each task has its individual decoder. The decoder for speech enhancement generates the enhanced speech $\hat{\mathbf{s}}$, while the decoder for VAD generates soft predictions $\hat{y}$.

The encoder is mainly a one-dimension convolutional layer with a kernel size of $L$ and stride $L/2$. It transforms the input noisy audio signal $\mathbf{x} \in \mathbb{R}^{1 \times T}$ to a feature map $\mathbf{W} \in \mathbb{R}^{N \times K}$, where $N$ and $K$ are the dimension and number of the feature vectors respectively. The TCN speech separation module estimates a mask $\mathbf{M} \in \mathbb{R}^{N \times K}$ from $\mathbf{W}$, and applies $\mathbf{M}$ to $\mathbf{W}$ by an element-wise multiplication, which gets the denoised feature map $\mathbf{D} \in \mathbb{R}^{N \times K}$, i.e. $\mathbf{D} = \mathbf{M}\odot \mathbf{W}$ where $\odot$ denotes the element-wise multiplication.

The decoders are two independent one-dimensional transposed convolution layers. Each of them conducts an opposite dimensional transform to the encoder. Both of the decoders take $\mathbf{D}$ as the input. They generate the estimated clean speech $\mathbf{\hat{s}} \in \mathbb{R}^{1 \times T}$ and VAD scores respectively. To generate probability-like soft decision scores for VAD, a sigmoid function is used to constrain the output of the VAD decoder between $0$ and $1$, which outputs $\mathbf{\hat{y}} = [\hat{y}_1, \ldots, \hat{y}_T]  \in {[0,1]}^{1\times T}$.

%

\subsection{Objective function and optimization}
\label{ssec:loss}

The end-to-end multi-task model uses the following joint loss:
\begin{equation}
	L=\lambda \ell_{\rm vad}+(1-\lambda)\ell_{\rm enhance}
\end{equation}
where $\ell_{\rm vad}$ and $\ell_{\rm enhance}$ are the loss components for VAD and speech enhancement respectively, and $\lambda \in (0,1)$ is a hyperparameter to balance the two components. We use the cross-entropy minimization as $\ell_{\rm vad}$. Because SI-SDR \cite{Luo2019} is frequently used as the optimization objective of end-to-end speech separation, a conventional thought of multitask learning is to optimize SI-SDR and cross-entropy jointly. However, the two decoders in this strategy are optimized independently, which do not benefit VAD and speech enhancement together. As we know, VAD and speech enhancement share many common properties. For example, the earliest ideal-binary-masking based speech enhancement can be regarded as VAD applied to each frequency band \cite{6317144}.

To benefit the advantages of VAD and speech enhancement together, we propose a new speech enhancement loss, named mSI-SDR, as $\ell_{\rm enhance}$ for the multi-task training. We present mSI-SDR in detail as follows.


mSI-SDR is revised from the conventional SI-SDR. SI-SDR is designed to solve the scale-dependent problem in the signal-to-distortion ratio \cite{Roux2019}:
\begin{equation}
	{\mbox{SI-SDR}}=10\log_{10}\dfrac{||\mathbf{e}_{target}||^{2}}{||\mathbf{e}_{res}||^{2}}=10\log_{10}\dfrac{||\alpha \mathbf{s}||^{2}}{||\alpha \mathbf{s}-\hat{\mathbf{s}}||^{2}}
\end{equation}
where $\mathbf{s}$ is the referenced signal, $\hat{\mathbf{s}}$ is the estimated signal, and $\alpha =\dfrac{\hat{\mathbf{s}}^{T}\mathbf{s}}{||\mathbf{s}||^{2}}$ denotes the scaling factor.

mSI-SDR introduces the VAD labels and predictions into SI-SDR:
\begin{equation}\label{eq:xx}
	\ell_{\rm enhance} = {\mbox{mSI-SDR}}=10\log_{10}\dfrac{||\beta \mathbf{s}||^{2}}{||\beta \mathbf{s}-\hat{\mathbf{s}}^{*}||^{2}}
\end{equation}
where
\begin{equation}\label{eq:yy}
	\hat{\mathbf{s}}^{*}=\hat{\mathbf{s}}+\hat{\mathbf{s}}\odot (\mathbf{y}+\hat{\mathbf{y}})
\end{equation}
 $\beta=\dfrac{\hat{\mathbf{s}}^{*T}\mathbf{s}}{||\mathbf{s}||^{2}}$, and $\mathbf{y}=[y_1,\ldots,y_T]$ is the ground-truth VAD label.
From \eqref{eq:xx}, we see that mSI-SDR takes the enhanced speech, clean speech, ground-truth VAD labels, and predicted VAD labels into consideration.

Equation \eqref{eq:yy} is important in benefitting VAD and SE together. It makes $\ell_{\rm enhance}$ focus on enhancing the voice active part of the signal. More importantly, when optimizing the joint loss function by gradient descent, the updating process of the VAD decoder depends on both $\ell_{\rm vad}$ and $\ell_{\rm enhance}$, which makes VAD use the two kinds of references sufficiently.

\section{Experiments}
\label{sec:experiments}

\subsection{Experimental setup}
\label{ssec:setup}

 Wall Street Journal (WSJ0) \cite{paul1992design} dataset was used as the source of clean speech. It contains 12776 utterances from 101 speakers for training, 1206 utterances from 10 speakers for validation, and 651 utterances from 8 speakers for evaluation. Only 20\% of the audio recordings is silence. To alleviate the class imbalanced problem, we added silent segments of 0.5 and 1 second to the front and end of each audio recording respectively. The noise source for training and development is a large-scale noise library containing over 20000 noise segments. The noise source for test is five unseen noises, where the bus, caffe, pedestrians, and street noise are from CHiME-3 dataset \cite{barker2015third}, and the babble noise is from the NOISEX-92 noise corpus \cite{varga1993assessment}. The SNR level of each noisy speech recording in the training and development sets was selected randomly from the range of $[-5, 5]$ dB. The SNR levels of the test sets were set to $-5$dB, 0dB, and 5dB respectively. The noise sources between training, development, and test do not overlap. All signals were resampled to 16 kHz. The ground-truth VAD labels were obtained by applying Ramirez VAD \cite{ramirez2005statistical} with human-defined smoothing rules to the clean speech. This method was proved to be reasonable for generating ground-truth labels \cite{zhang2015boosting,Xu2019,Jung2018}.

 \begin{table*}[t]
	\caption{Performance of the Multi-mSS, Multi-SS, and VAD-only models for VAD.}
	\label{vad}
	\centering
\scalebox{0.8}{
	\begin{tabular}{c|c|c|c|c|c|c|c|c|c|c|c|c|ccc}
		\hline
		\multirow{2}{*}{\textbf{Noise}} & \multirow{2}{*}{\textbf{\begin{tabular}[c]{@{}c@{}}SNR\\ (dB)\end{tabular}}} & \multicolumn{3}{c|}{\textbf{AUC(\%)}}                                                                                                                            & \multicolumn{3}{c|}{\textbf{EER(\%)}}                                                                                                                            & \multirow{2}{*}{\textbf{Noise}} & \multirow{2}{*}{\textbf{\begin{tabular}[c]{@{}c@{}}SNR\\ (dB)\end{tabular}}} & \multicolumn{3}{c|}{\textbf{AUC(\%)}}                                                                                                                            & \multicolumn{3}{c}{\textbf{EER(\%)}}                                                                                                                                                                       \\ \cline{3-8} \cline{11-16}
		&                                                                              & \begin{tabular}[c]{@{}c@{}}Multi-\\ mSS\end{tabular} & \begin{tabular}[c]{@{}c@{}}Multi-\\ SS\end{tabular} & \begin{tabular}[c]{@{}c@{}}VAD-\\ only\end{tabular} & \begin{tabular}[c]{@{}c@{}}Multi-\\ mSS\end{tabular} & \begin{tabular}[c]{@{}c@{}}Multi-\\ SS\end{tabular} & \begin{tabular}[c]{@{}c@{}}VAD-\\ only\end{tabular} &                                 &                                                                              & \begin{tabular}[c]{@{}c@{}}Multi-\\ mSS\end{tabular} & \begin{tabular}[c]{@{}c@{}}Multi-\\ SS\end{tabular} & \begin{tabular}[c]{@{}c@{}}VAD-\\ only\end{tabular} & \multicolumn{1}{c|}{\begin{tabular}[c]{@{}c@{}}Multi-\\ mSS\end{tabular}} & \multicolumn{1}{c|}{\begin{tabular}[c]{@{}c@{}}Multi-\\ SS\end{tabular}} & \begin{tabular}[c]{@{}c@{}}VAD-\\ only\end{tabular} \\ \hline
		\multirow{3}{*}{Babble}         & -5                                                                           & \textbf{98.4}                                        & 97.7                                                & 93.9                                                & \textbf{4.68}                                        & 5.63                                                & 11.65                                               & \multirow{3}{*}{Pedestrains}    & -5                                                                           & \textbf{98.6}                                        & 98.5                                                & 96.9                                                & \multicolumn{1}{c|}{\textbf{4.46}}                                        & \multicolumn{1}{c|}{4.58}                                                & 7.93                                                \\
		& 0                                                                            & \textbf{99.6}                                        & \textbf{99.6}                                       & 98.4                                                & \textbf{2.19}                                        & 2.22                                                & 4.83                                                &                                 & 0                                                                            & \textbf{99.5}                                        & \textbf{99.5}                                       & 99.3                                                & \multicolumn{1}{c|}{\textbf{2.44}}                                        & \multicolumn{1}{c|}{2.59}                                                & 3.23                                                \\
		& 5                                                                            & \textbf{99.7}                                        & \textbf{99.7}                                       & 99.4                                                & \textbf{1.72}                                        & 1.89                                                & 2.74                                                &                                 & 5                                                                            & \textbf{99.7}                                        & 99.6                                                & 99.6                                                & \multicolumn{1}{c|}{\textbf{1.87}}                                        & \multicolumn{1}{c|}{2.05}                                                & 2.34                                                \\ \hline
		\multirow{3}{*}{Bus}            & -5                                                                           & \textbf{99.5}                                        & 99.4                                                & 99.4                                                & \textbf{2.24}                                        & 2.49                                                & 2.85                                                & \multirow{3}{*}{Street}         & -5                                                                           & \textbf{99.3}                                        & \textbf{99.3}                                       & 98.9                                                & \multicolumn{1}{c|}{\textbf{2.80}}                                        & \multicolumn{1}{c|}{2.82}                                                & 4.01                                                \\
		& 0                                                                            & \textbf{99.7}                                        & 99.6                                                & 99.6                                                & \textbf{1.79}                                        & 2.00                                                & 2.29                                                &                                 & 0                                                                            & \textbf{99.7}                                        & 99.6                                                & 99.5                                                & \multicolumn{1}{c|}{\textbf{1.93}}                                        & \multicolumn{1}{c|}{2.04}                                                & 2.45                                                \\
		& 5                                                                            & \textbf{99.7}                                        & \textbf{99.7}                                       & 99.7                                                & \textbf{1.45}                                        & 1.66                                                & 1.83                                                &                                 & 5                                                                            & \textbf{99.7}                                        & \textbf{99.7}                                       & 99.6                                                & \multicolumn{1}{c|}{\textbf{1.62}}                                        & \multicolumn{1}{c|}{1.75}                                                & 2.14                                                \\ \hline
		\multirow{3}{*}{Caffe}          & -5                                                                           & \textbf{98.9}                                        & 98.7                                                & 97.1                                                & \textbf{4.00}                                        & 4.27                                                & 7.47                                                & \multirow{3}{*}{Average}        & -5                                                                           & \textbf{99.0}                                        & 98.9                                                & 97.5                                                & \multicolumn{1}{c|}{\textbf{3.59}}                                        & \multicolumn{1}{c|}{3.80}                                                & 6.67                                                \\
		& 0                                                                            & \textbf{99.6}                                        & 99.5                                                & 99.3                                                & \textbf{2.31}                                        & 2.48                                                & 3.25                                                &                                 & 0                                                                            & \textbf{99.6}                                        & 99.5                                                & 99.2                                                & \multicolumn{1}{c|}{\textbf{2.18}}                                        & \multicolumn{1}{c|}{2.34}                                                & 3.11                                                \\
		& 5                                                                            & \textbf{99.7}                                        & 99.6                                                & 99.6                                                & \textbf{1.77}                                        & 2.00                                                & 2.31                                                &                                 & 5                                                                            & \textbf{99.7}                                        & \textbf{99.7}                                       & 99.6                                                & \multicolumn{1}{c|}{\textbf{1.68}}                                        & \multicolumn{1}{c|}{1.86}                                                & 2.20                                                \\ \hline
	\end{tabular}
}
\end{table*}

We denote the proposed method as the \textit{multi-task model with mSI-SDR loss} (Multi-mSS).
For the model training, each training audio recording was cropped into several 4-second segments. The mini-batch size was set to 8. The Adam optimizer \cite{kingma2014adam} was used. The initial learning rate was set to $1e^{-3}$ and will be halved if the performance on the validation set has no improvement in 3 consecutive epochs. The minimum learning rate was set to $1e^{-8}$. The weight decay was set to $1e^{-5}$. The training was stopped if not performance improvement was observed in 6 consecutive epochs. The specific parameter setting of the end-to-end network follow the default setting of Conv-Tasnet \cite{Luo2019} with $L=32$.

To compare with Multi-mSS, we trained a \textit{multi-task model with SI-SDR loss} (Multi-SS) and a VAD-only single-task model denoted as \textit{VAD-only model}. Multi-SS has exactly the same network structure as Multi-mSS. The objective of its SE decoder was set to SI-SDR.
The VAD-only model removes the SE decoder and uses the VAD loss $\ell_{vad}$ as the optimization objective. We used the receiver-operating-characteristic (ROC) curve, area under the ROC curve (AUC), and equal error rate (EER) as the evaluation metrics for VAD. We took the signal of every 10ms as an observation for the calculation of AUC and EER. We used the perceptual evaluation of speech quality (PESQ), short-time objective intelligibility (STOI) \cite{taal2010short}, and scale-invariant source-to-distortion ratio (SI-SDR) \cite{Roux2019} as the evaluation metrics for speech enhancement.


\subsection{Results}

\label{ssec:result1}

\begin{figure}[t]
	\centering
	\centerline{\includegraphics[width=5cm]{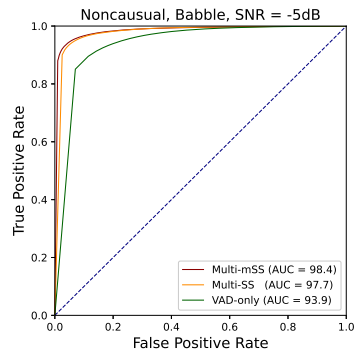}}
	\caption{ROC curve comparison in the babble noise at $-5$ dB.}
	\label{fig:roc}
\end{figure}

\begin{figure}[t]
	
	\begin{minipage}[b]{0.48\linewidth}
		\centering
		\centerline{\includegraphics[width=4.8cm]{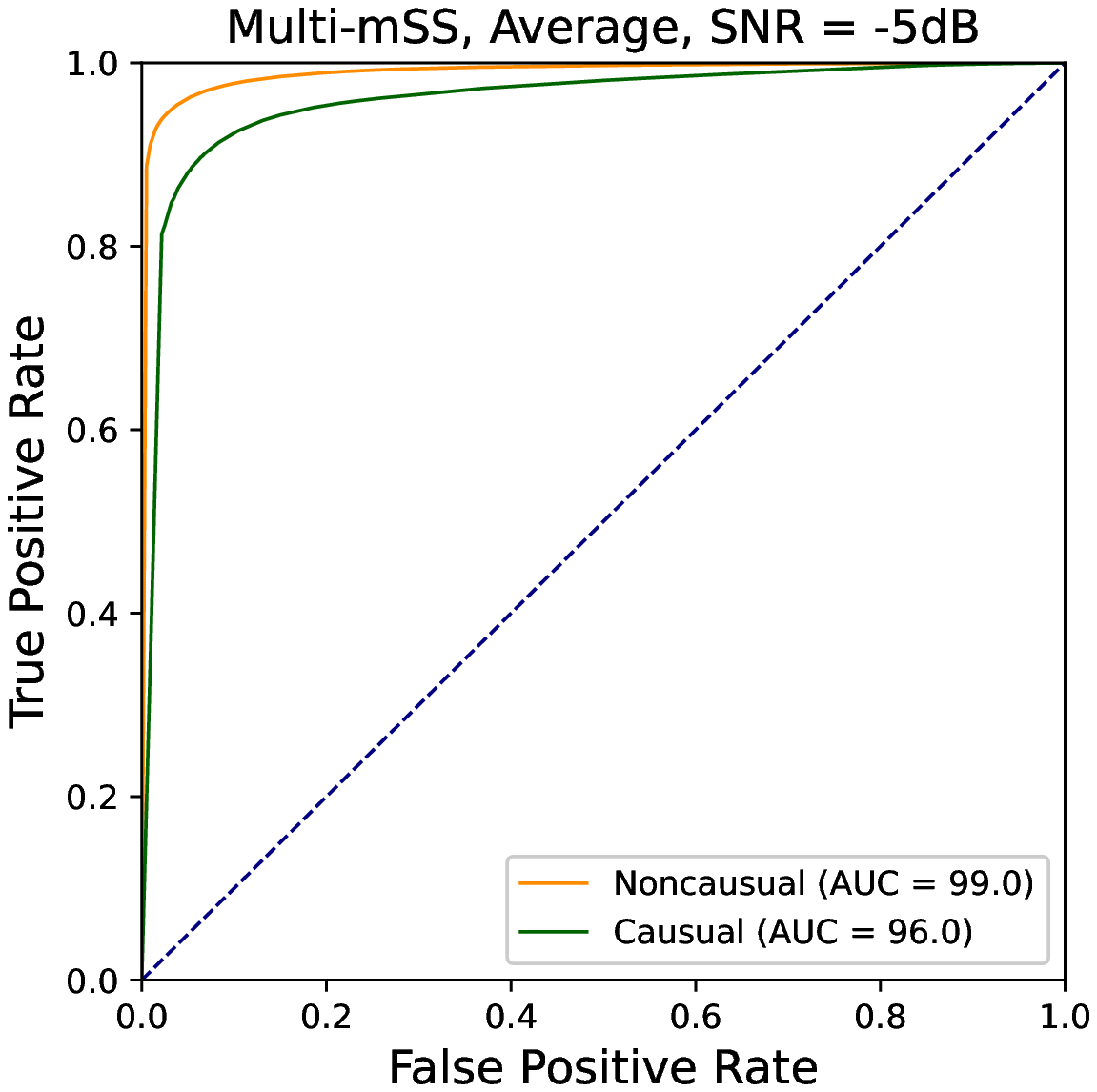}}
		\centerline{(a)}\medskip
	\end{minipage}
	\hfill
	\begin{minipage}[b]{0.48\linewidth}
		\centering
		\centerline{\includegraphics[width=4.8cm]{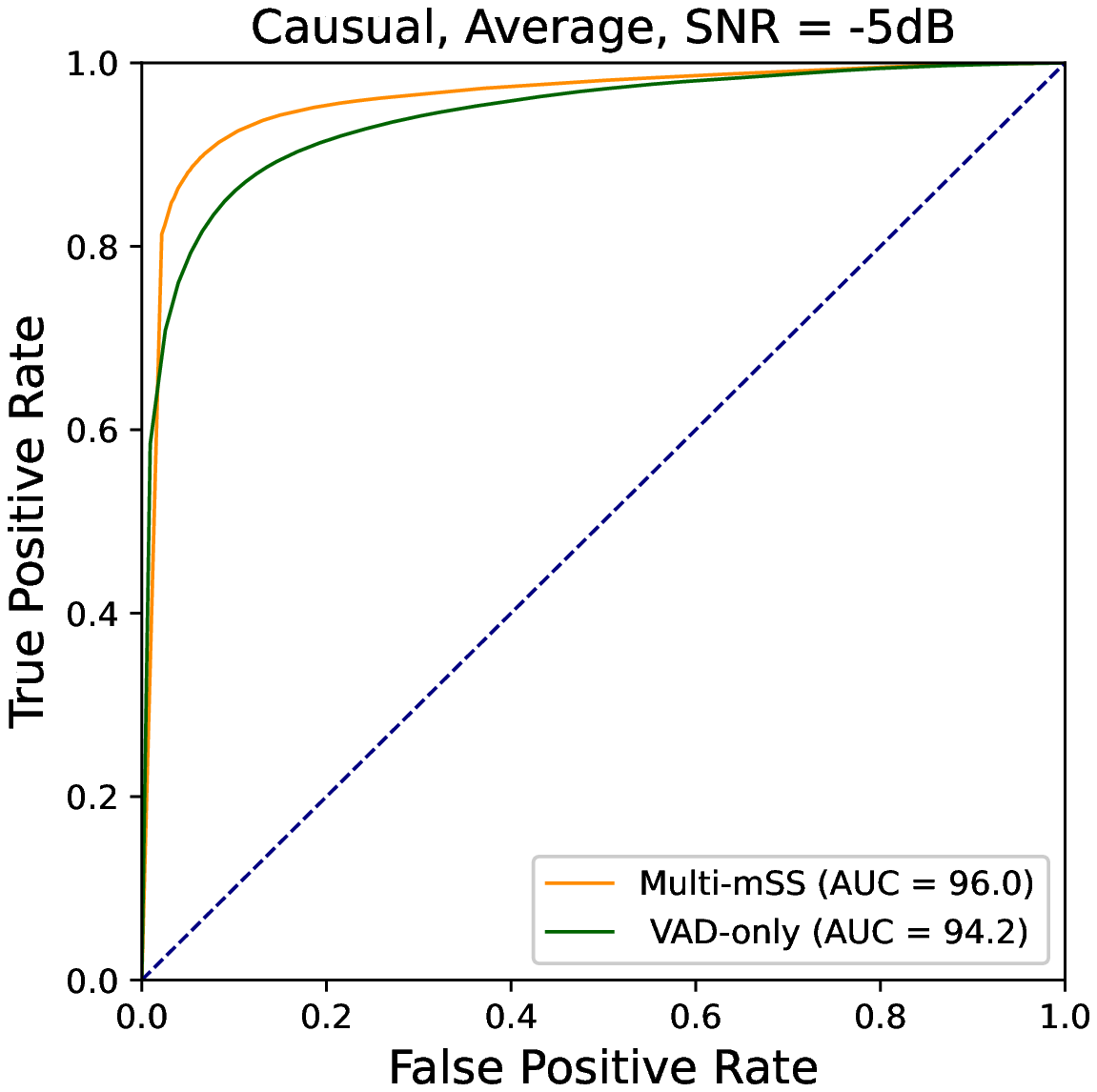}}
		\centerline{(b)}\medskip
	\end{minipage}
	\caption{ROC curve comparison in causual configurations.}
	\label{fig:roc_cau}
\end{figure}


\textbf{Comparison between Multi-mSS and the VAD-only model:}
The comparison result between the proposed Multi-mSS and the VAD-only model is shown in Table \ref{vad}. From the table, we see that Multi-mSS outperforms the VAD-only model in all noise environments and SNR conditions in terms of both AUC and EER. The relative performance improvement is enlarged when the SNR level becomes low. For example, Multi-mSS provides a relative AUC improvement of 73.77\% over the VAD-only model, and a relative EER reduction of 59.83\% over the latter in the babble noise at $-5$ dB. When the SNR is increased to 5 dB, the relative improvement is reduced to 50.00\% and 37.23\% respectively.

From the table, we also notice that the advantage of Multi-mSS is obvious in difficult noisy environments. Specifically,
The relative EER reduction in the babble, caffe and pedestrains environments is 55.38\%, 38.02\% and 35.11\% respectively. In contrast, the relative EER reduction in the bus and street environments is only 21.12\% and 26.13\%. One can see that the babble, caffe and pedestrains environments are  speech-shaped ones, which have similar distributions with the targeted speech. 


Although our goal is to improve the performance of VAD, we also list the comparison of Multi-mSS and the SE-only single-task model (denoted as \textit{SE-only model}) on SE performance here as a reference. The result in Table \ref{se} shows that the performance of the speech enhancement task was not greatly affected.

\begin{table}[t]
	\caption{Average performance of the Multi-mSS, Multi-SS, and SE-only models for speech enhancement.}
	\label{se}
	\centering
	\scalebox{0.75}{
		\begin{tabular}{c|c|ccc}
			\hline
			\multirow{2}{*}{\textbf{Metrics}} & \multirow{2}{*}{\textbf{Model}} & \multicolumn{3}{c}{\textbf{SNR(dB)}} \\ \cline{3-5}
			&                                 & -5dB       & 0dB        & 5dB        \\ \hline
			\multirow{3}{*}{PESQ}             & Multi-mSS                         & 2.422      & 2.848      & 3.151      \\
			& Multi-SS                           & 2.404      & 2.856      & 3.168      \\
			& SE-only                         & 2.457      & 2.906      & 3.224      \\ \hline
			\multirow{3}{*}{STOI}             & Multi-mSS                          & 0.898      & 0.950      & 0.972      \\
			& Multi-SS                          & 0.897      & 0.950      & 0.973      \\
			& SE-only                         & 0.898      & 0.950      & 0.973      \\ \hline
			\multirow{3}{*}{SI-SDR}           & Multi-mSS                          & 9.705      & 13.176     & 16.105     \\
			& Multi-SS                          & 9.829      & 13.529     & 16.674     \\
			& SE-only                         & 9.873      & 13.577     & 16.721     \\ \hline
		\end{tabular}
	}
\end{table}

\textbf{Comparison between Multi-mSS and Multi-SS:}
Table \ref{vad} also shows the comparison result between Multi-mSS and Multi-SS. From the table, we see that Muli-mSS produces at least comparable performance to Muli-SS in all environments. Particularly, Multi-mSS provides a relative AUC improvement of 30.43\% and a relative EER reduction of 16.87\% over Multi-SS in the most difficult environment---babble noise at $-5$ dB, where the ROC curves of the three comparison methods are further drawn in Fig. \ref{fig:roc}.

\textbf{Comparison with causual configurations:}
We also evaluated the comparison methods with the same causal configurations as \cite{Luo2019}. Specifically, we first replaced the global layer normalization with cumulative layer normalization, and then used causal dilated convolution in TCN. This makes the comparison methods work in real time with a minimum delay of about 2ms. Fig. \ref{fig:roc_cau} shows the average ROC curves of the comparison methods over all 5 noisy conditions at $-5$ dB. From Fig. \ref{fig:roc_cau}a, we see that the causal Multi-mSS does not suffer much performance degradation from the noncausal Multi-mSS. From Fig. \ref{fig:roc_cau}b, we see that the causal Multi-mSS outperforms the causal VAD-only model significantly, which is consistent to the conclusion in the noncausal configurations. 

%


\section{conclusions}
\label{sec:conclusions}

In this paper, we have proposed an end-to-end multi-task model with a novel loss funtion named VAD-masked scale-invariant source-to-distortion ratio (mSI-SDR) to increase robustness of the VAD system in low SNR environments. mSI-SDR takes the VAD information into the optimization of the SE decoder, which makes the two tasks jointly optimized not only at the encoder and separation networks, but also at the objective level.An additional merit is that it theoretically satisfies real-time applications. Experimental results show that the proposed method outperforms the VAD-only model in all noise conditions, especially the low SNR environments and that with much human voice interference. Moreover, mSI-SDR yields better performance than SI-SDR in the multi-task setting. In the future, we will evaluate the proposed method in more complicated scenarios and compare it with the state-of-the-art VAD in the system level \cite{kimtool}.

\vfill\pagebreak

%

\small
\bibliographystyle{IEEEbib}
\bibliography{strings,refs}

\begin{thebibliography}{10}

\bibitem{zhang2012deep}
Xiao-Lei Zhang and Ji~Wu,
\newblock ``Deep belief networks based voice activity detection,''
\newblock {\em IEEE/ACM TASLP}, vol. 21, no. 4, pp. 697--710, 2012.

\bibitem{hughes2013recurrent}
Thad Hughes and Keir Mierle,
\newblock ``Recurrent neural networks for voice activity detection,''
\newblock in {\em in ICASSP, 2013}. IEEE, 2013, pp. 7378--7382.

\bibitem{thomas2014analyzing}
Samuel Thomas, Sriram Ganapathy, George Saon, and Hagen Soltau,
\newblock ``Analyzing convolutional neural networks for speech activity
  detection in mismatched acoustic conditions,''
\newblock in {\em ICASSP, 2014}. IEEE, 2014, pp. 2519--2523.

\bibitem{zhang2015boosting}
Xiao-Lei Zhang and DeLiang Wang,
\newblock ``Boosting contextual information for deep neural network based voice
  activity detection,''
\newblock {\em IEEE/ACM TASLP}, vol. 24, no. 2, pp. 252--264, 2015.

\bibitem{kim2018multi}
Jaeseok Kim, Heejin Choi, Jinuk Park, Juntae Kim, and Minsoo Hahn,
\newblock ``Voice activity detection based on multi-dilated convolutional
  neural network,''
\newblock in {\em ICMSCE, 2018}, 2018, p. 98–102.

\bibitem{KIM2018}
J.~{Kim} and M.~{Hahn},
\newblock ``Voice activity detection using an adaptive context attention
  model,''
\newblock {\em IEEE Signal Processing Letters}, vol. 25, no. 8, pp. 1181--1185,
  2018.

\bibitem{chang2018}
S.~{Chang}, B.~{Li}, G.~{Simko}, T.~N. {Sainath}, A.~{Tripathi}, A.~{van den
  Oord}, and O.~{Vinyals},
\newblock ``Temporal modeling using dilated convolution and gating for
  voice-activity-detection,''
\newblock in {\em ICASSP, 2018}, 2018, pp. 5549--5553.

\bibitem{Fernando2020}
Tharindu Fernando, Sridha Sridharan, Mitchell McLaren, Darshana Priyasad, Simon
  Denman, and Clinton Fookes,
\newblock ``{Temporarily-Aware Context Modeling Using Generative Adversarial
  Networks for Speech Activity Detection},''
\newblock {\em IEEE/ACM TASLP}, vol. 28, pp. 1159--1169, 2020.

\bibitem{Zazo2016}
Ruben Zazo, Tara~N. Sainath, Gabor Simko, and Carolina Parada,
\newblock ``Feature learning with raw-waveform cldnns for voice activity
  detection,''
\newblock in {\em Interspeech 2016}, 2016.

\bibitem{Ariav2019}
Ido Ariav and Israel Cohen,
\newblock ``{An End-to-End Multimodal Voice Activity Detection Using WaveNet
  Encoder and Residual Networks},''
\newblock {\em IEEE Journal on Selected Topics in Signal Processing}, vol. 13,
  no. 2, pp. 265--274, 2019.

\bibitem{Yu2020}
Cheng Yu, Kuo-Hsuan Hung, I-Fan Lin, Szu-Wei Fu, Yu~Tsao, and Jeih weih Hung,
\newblock ``Waveform-based voice activity detection exploiting fully
  convolutional networks with multi-branched encoders,''
\newblock {\em arXiv preprint arXiv:2006.11139}, 2020.

\bibitem{zhang2013denoising}
Xiao-Lei Zhang and Ji~Wu,
\newblock ``Denoising deep neural networks based voice activity detection,''
\newblock in {\em ICASSP, 2013}. IEEE, 2013, pp. 853--857.

\bibitem{Wang2015}
Qing Wang, Jun Du, Xiao Bao, Zi~Rui Wang, Li~Rong Dai, and Chin~Hui Lee,
\newblock ``{A universal VAD based on jointly trained deep neural networks},''
\newblock {\em in Interspeech, 2015}, vol. 2015-Janua, pp. 2282--2286, 2015.

\bibitem{Lin2019}
Ruixi Lin, Charles Costello, Charles Jankowski, and Vishwas Mruthyunjaya,
\newblock ``{Optimizing voice activity detection for noisy conditions},''
\newblock {\em in Interspeech, 2019}, vol. 2019-September, pp. 2030--2034,
  2019.

\bibitem{Xu2019}
Tianjiao Xu, Hui Zhang, and Xueliang Zhang,
\newblock ``{Joint training ResCNN-based voice activity detection with speech
  enhancement},''
\newblock {\em in APSIPA ASC, 2019}, pp. 1157--1162, 2019.

\bibitem{Lee2020}
Geon~Woo Lee and Hong~Kook Kim,
\newblock ``{Multi-task learning U-Net for single-channel speech enhancement
  and mask-based voice activity detection},''
\newblock {\em Applied Sciences (Switzerland)}, vol. 10, no. 9, 2020.

\bibitem{Jung2018}
Youngmoon Jung, Younggwan Kim, Yeunju Choi, and Hoirin Kim,
\newblock ``{Joint learning using denoising variational autoencoders for voice
  activity detection},''
\newblock {\em in Interspeech, 2018}, vol. 2018-Septe, no. January 2019, pp.
  1210--1214, 2018.

\bibitem{Zhuang2017}
Yimeng Zhuang, Sibo Tong, Maofan Yin, Yanmin Qian, and Kai Yu,
\newblock ``{Multi-task joint-learning for robust voice activity detection},''
\newblock {\em in ISCSLP, 2016}, , no. 1, 2017.

\bibitem{Luo2019}
Yi~Luo and Nima Mesgarani,
\newblock ``{Conv-TasNet: Surpassing Ideal Time-Frequency Magnitude Masking for
  Speech Separation},''
\newblock {\em IEEE/ACM TASLP}, vol. 27, no. 8, pp. 1256--1266, 2019.

\bibitem{6317144}
Y.~{Wang}, K.~{Han}, and D.~{Wang},
\newblock ``Exploring monaural features for classification-based speech
  segregation,''
\newblock {\em IEEE/ACM TASLP}, vol. 21, no. 2, pp. 270--279, 2013.

\bibitem{Roux2019}
Jonathan~Le Roux, Scott Wisdom, Hakan Erdogan, and John~R. Hershey,
\newblock ``{SDR - Half-baked or Well Done?},''
\newblock {\em in ICASSP, 2019}, vol. 2019-May, pp. 626--630, 2019.

\bibitem{paul1992design}
Douglas~B Paul and Janet Baker,
\newblock ``The design for the wall street journal-based csr corpus,''
\newblock in {\em Speech and Natural Language: Proceedings of a Workshop Held
  at Harriman, New York, February 23-26, 1992}, 1992.

\bibitem{barker2015third}
Jon Barker, Ricard Marxer, Emmanuel Vincent, and Shinji Watanabe,
\newblock ``The third ‘chime’speech separation and recognition challenge:
  Dataset, task and baselines,''
\newblock in {\em 2015 IEEE Workshop on Automatic Speech Recognition and
  Understanding (ASRU)}. IEEE, 2015, pp. 504--511.

\bibitem{varga1993assessment}
Andrew Varga and Herman~JM Steeneken,
\newblock ``Assessment for automatic speech recognition: Ii. noisex-92: A
  database and an experiment to study the effect of additive noise on speech
  recognition systems,''
\newblock {\em Speech communication}, vol. 12, no. 3, pp. 247--251, 1993.

\bibitem{ramirez2005statistical}
Javier Ram{\'\i}rez, Jos{\'e}~C Segura, Carmen Ben{\'\i}tez, Luz Garc{\'\i}a,
  and Antonio Rubio,
\newblock ``Statistical voice activity detection using a multiple observation
  likelihood ratio test,''
\newblock {\em IEEE Signal Processing Letters}, vol. 12, no. 10, pp. 689--692,
  2005.

\bibitem{kingma2014adam}
Diederik~P Kingma and Jimmy Ba,
\newblock ``Adam: A method for stochastic optimization,''
\newblock {\em arXiv preprint arXiv:1412.6980}, 2014.

\bibitem{taal2010short}
Cees~H Taal, Richard~C Hendriks, Richard Heusdens, and Jesper Jensen,
\newblock ``A short-time objective intelligibility measure for time-frequency
  weighted noisy speech,''
\newblock in {\em ICASSP, 2010}. IEEE, 2010, pp. 4214--4217.

\bibitem{kimtool}
Juntae Kim,
\newblock ``Voice activity detection toolkit website,''
  \url{https://github.com/jtkim-kaist/VAD}, 2018.

\end{thebibliography}

\end{document}